\documentstyle[epsf]{mn}

\def\gsim{\mathrel{\raise0.35ex\hbox{$\scriptstyle >$}\kern-0.6em 
\lower0.40ex\hbox{{$\scriptstyle \sim$}}}}
\def\lsim{\mathrel{\raise0.35ex\hbox{$\scriptstyle <$}\kern-0.6em 
\lower0.40ex\hbox{{$\scriptstyle \sim$}}}}

\title{Probing the evolution of early-type galaxies using multi-colour
number counts and redshift distributions}
\author[Nakata et al.]
{
Fumiaki Nakata$^1$, Kazuhiro Shimasaku$^{1,2}$, Mamoru Doi$^{1,2}$,
Nobunari Kashikawa$^3$, \\ \\
\LARGE{Wataru Kawasaki$^1$, Yutaka Komiyama$^1$,
Sadanori Okamura$^{1,2}$, Maki Sekiguchi$^4$,} \\ \\
\LARGE{Masafumi Yagi$^3$ and Naoki Yasuda$^3$} \\ \\
$^1$ Department of Astronomy, School of Science, University of Tokyo, Japan \\
$^2$ Research Center for the Early Universe, University of Tokyo, Japan \\
$^3$ National Astronomical Observatory of Japan \\
$^4$ Institute for Cosmic Ray Research, University of Tokyo, Japan
}

\begin{document}

\maketitle

\begin{abstract}

We investigate pure luminosity evolution models for
early-type (elliptical and S0) galaxies (i.e., no number
density change or morphology transition), and examine whether
these models are consistent with observed number counts in the $B$, $I$ and $
K$ bands and redshift distributions of two samples of faint galaxies selected
in the $I$ and $K$ bands. The models are characterized by the star formation
time scale $\tau_{\rm SF}$ and the time $t_{\rm gw}$ when galactic wind blows
in addition to several other conventional parameters. We find the single-burst
model ($\tau_{\rm SF}=0.1$ Gyr and $t_{\rm gw}=0.353$ Gyr), which is known
to reproduce the photometric properties of early-type galaxies in
clusters, is inconsistent with redshift distributions of early-type
galaxies in the field environment due to overpredictions of galaxies at $
z\gsim1.4$ even with strong extinction which is at work until $t_{\rm gw}
$. In order for dust extinction to be more effective, we change $
\tau_{\rm SF}$ and $t_{\rm gw}$ as free parameters, and find that models
with $\tau_{\rm SF}\gsim0.5$ Gyr and $t_{\rm gw}>1.0$ Gyr can be made
consistent with both the observed redshift distributions and number
counts, if we introduce strong extinction $(E(B-V)\geq1$ as a peak
value). These results suggest that early-type galaxies in the field
environment do not have the same evolutionary history as described by the
single-burst model.

\end{abstract}


\section{INTRODUCTION}\label{sec:intro}

How early-type galaxies were formed and
evolved is a key issue in extragalactic astronomy which remains
controversial. Evolution of early-type galaxies in clusters seems to be well
expressed by the so-called single-burst model in which galaxies experience a
star burst at the initial phase of their formation and then evolve passively
without any subsequent star formation. For example, Ellis et al. (1997) found
that the scatter in the colour-magnitude relation of early-type galaxies is
quite small for $z\sim0.5$ clusters, indicating that the major star
formation ended at $z\gsim3$ (See also
Stanford, Eisenhardt \& Dickinson 1998). Kodama et al. (1998) investigated
the slope and zero-point of the colour-magnitude relation in 17 distant
clusters with redshift $0.31<z<1.27$, and constrained the epoch of major
star formation to $z>2-4$. However, it is fairly controversial whether the
single-burst model holds for early-type galaxies in the field
environment. Abraham et al. (1999) analysed the spatially resolved colours
of galaxies detected in the Hubble Deep Field (HDF; Williams et al. 1996) and
argued that 40 \% of early-type galaxies have experienced the major star
formation at $z<1$. Kodama, Bower \& Bell (1999) obtained similar results
on the basis of colour-magnitude diagram of HDF early-type
galaxies. Franceschini et al. (1998) compared model spectra with a
seven-colour broad band photometry of HDF early-type galaxies, and concluded
that the major episodes of star formation building up typical $M^\ast
$ galaxies took place during $1<z<4$ for the deceleration parameter $q_0=0.5
$ ($1<z<3$ for $q_0=0.15$). On the other hand, Bernardi et
al. (1998) showed, on the basis of the MgII-$\sigma$ relation, that nearby
early-type galaxies in clusters and in the field environment consist of the
same stellar population, and claimed that the bulk of stars in the early-type
galaxies had to form at $z\gsim3$ in the field environment as well as in
clusters.

In this paper, we first examine whether the single burst model reproduces
both the observed number counts in the $B$, $I$ and $K$ bands and redshift
distributions of two samples of HDF galaxies selected in the $I$ and $K
$ bands. We will find that the single-burst model overpredicts galaxies
at $z\gsim1.4$ in redshift distribution even if the effect of dust extinction
is taken into account during star formation. Following this result, we then
introduce models which have longer periods of star formation than the
single-burst model so that dust extinction work longer. We compare these
models with the observations, and find that with a certain combination of
parameters there are models whose predictions are consistent with the
observations. Our study can be regarded as an extension of the work by
Franceschini et al. (1998) in which they compared the single burst model
and a model having a longer star formation period with $K$-band number
counts and redshift distribution. We limit our discussion to so-called
pure luminosity evolution
models, because there is non-controversial evidence for number
evolution or morphology evolution for early-type galaxies.

The structure of this paper is as follows. In \S2, we describe the data used
for our analysis, and in \S3 we show how the local luminosity function is
normalised. We compare the observed number counts and redshift distributions
with models in \S4. Our conclusions are given in \S5. The cosmological
parameters we have chosen to use throughout this paper are $
(\Omega_0,$ $\lambda_0,$ $t_0)=(0.1,$ $0,$ $15$ Gyr), unless otherwise
stated, where $\Omega_0$ is the density parameter, $\lambda_0$ is the
cosmological constant, and $t_0$ is the present age of the universe (for this
choice $h=0.59$ where $H_0\equiv100 h$ km s$^{-1}$ Mpc$^{-1}
$). Note, however, that we
have also examined models for two other cosmologies, $(\Omega_0,$ $\lambda_0,
$ $t_0)=(1.0,$ $0,$ $13$ Gyr) and (0.1, 0.9, 15 Gyr), and found that main
results do not change.

\section{DESCRIPTION OF THE DATA}

We use $B$-, $I$- and $K$-band number counts and redshift
distributions of early-type galaxies. The $B$-band number count in the
bright magnitude range is obtained using our mosaic CCD
camera (Kashikawa et al. 1995) attached to the Las
Campanas 1-m Telescope in 1995 October and to the 4.2-m William Herschel
Telescope in 1996 April (4.30 deg$^2$ in total; hereafter MCCD data). The
$I$- and $K$- band number counts in the bright magnitude range are taken from
Huang, Cowie \& Luppino (1998). Number counts in the faint magnitude
range are taken from work based on the Hubble Space
Telescope (HST) observations (Driver, Windhorst \& Griffiths 1995; Driver et
al. 1995; Glazebrook et al. 1995; Casertano et al. 1995; Abraham et
al. 1996a,b; Odewahn et al. 1996; Driver et al. 1998; Franceschini et
al. 1998).

In deriving number counts of early-type galaxies, morphological
classification is needed. MCCD data are classified into early-type and
late-type galaxies using the central concentration and the average surface
brightness of galaxies (Doi, Fukugita \& Okamura 1993). A part of $I$-band HST
data (Abraham et al. 1996a,b) is classified using the central concentration
and the asymmetry of galaxies. Morphological classification of other data is
mostly based on eye-ball inspection, though some authors used profile
fitting or neural network as well (Casertano et al. 1995; Odewahn
et al. 1996; Driver et al. 1998; Franceschini et al. 1998). Morphologically
classified $K$-band counts are available from only two sources (Huang et
al. 1998; Franceschini et al. 1998).

We use redshift distributions of the HDF galaxies selected in the $I
$ and $K$ bands. The $I$-band sample is
taken from Driver et al. (1998). There are 47 galaxies in $22<I<26$.
Only 12 out of the 47 galaxies have spectroscopic redshifts, and
photometric redshifts (Fern\'{a}ndez-Soto, Lanzetta \& Yahil 1999) are
used for the remaining 35. The $K$-band sample is taken from
Franceschini et al. (1998). There are 34 galaxies
in $16.15<K<20.15$. Out of the 34 galaxies 14 have
spectroscopic redshifts and 20 are assigned photometric
redshifts (Franceschini et al. 1998).

\section{NORMALISATION OF LOCAL LUMINOSITY FUNCTION}

To predict number counts and redshift distributions, the local luminosity
function is needed. Recently, the local luminosity function of early-type
galaxies has been derived using large redshift surveys. We consider two
recent surveys, SSRS2 (Marzke et al. 1998) and 2dF (Colless 1998). The
values for $M^\ast$ and $\alpha$ are different between them (($
\alpha,$ $M^\ast-5\log h)=(-1.00,$ $-19.37)$ for SSRS2 and $(-0.486,$ $
-19.46)$ for 2dF). We adopt the values given in SSRS2, because the 2dF
results are preliminary. However, we confirm that the differences in $
M^\ast$ and $\alpha$ between SSRS2 and 2dF do not significantly affect our
results. The measured values for normalisation factor $\phi^\ast$ are
different by about a factor of 2 between SSRS2 (0.0086 $h^3$Mpc$^{-3}$) and
2dF (0.0044 $h^3$Mpc$^{-3}$). We do not take these $\phi^\ast
$ value. Instead, we determine $\phi^\ast$ so that the predicted counts
at $17.5<B<20$ match our MCCD data. The resulting value $\phi^\ast=0.0067
$ $h^3$Mpc$^{-3}$, which is $\sim$1.5 times lower than the SSRS2 value, is
adopted.

\section{ANALYSIS}

\subsection{Single burst model of early-type galaxies}

\begin{figure}
\begin{center}
  \epsfxsize 0.9\hsize
  \epsffile{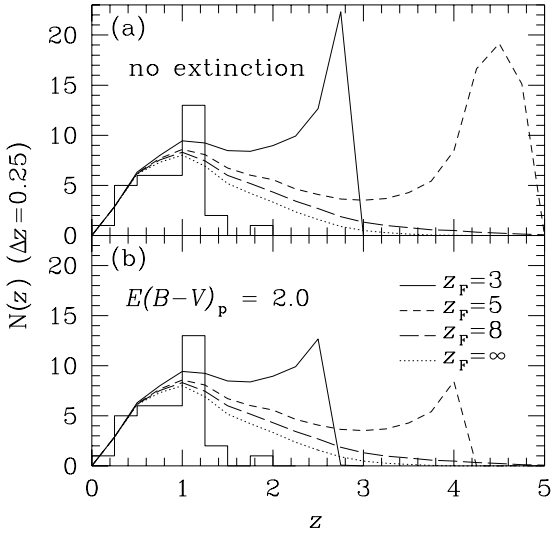}
\end{center}
\vspace{-0.8cm}
\caption{Redshift distribution for the $K$-band sample ($16.15<K<20.15
$) (histograms). Predictions of the single-burst models with $
z_{\rm F}=3,$ 5, 8 and $\infty$ are shown by lines. Panels (a) and (b) are
for no extinction and for the strongest extinction we consider, respectively.
\label{fig:nz_singleburst}}
\end{figure}

We assume pure luminosity evolution for early-type galaxies without the
evolution of number or morphology. We compute galaxy
evolution using the spectral synthesis code by
Kodama \& Arimoto (1997) (hereafter KA97). The star formation rate of
galaxies $\psi$ is expressed as

\begin{eqnarray}
\left\{
  \begin{array}{ll}
  \psi=\frac{1}{\tau_{\rm SF}}M_{\rm gas} & (t<t_{\rm gw}) \\
  \psi\simeq0 & (t\ge t_{\rm gw})
  \end{array}
\right.
\end{eqnarray}

\noindent
where $\tau_{\rm SF}
$ is the star formation time scale, $t_{\rm gw}$ is the time when galactic
wind blows, and $M_{\rm gas}$ is the gas mass. Luminosity evolution of
early-type galaxies, which we are concerned with, is characterized mostly by
two parameters, $\tau_{\rm SF}$ and $t_{\rm gw}$. KA97 found that the observed
colour-magnitude relation of early-type galaxies in clusters is well
reproduced if they adopt $\tau_{\rm SF}=0.1$ Gyr, $t_{\rm gw}<0.353
$ Gyr, and the galaxy formation redshift $z_{\rm F}=4.5$ ($q_0=0.5,$ $h=0.5
$). We limit our discussion to models with $z_{\rm F}\geq3$, because
early-type galaxies in clusters are suggested to have $z_{\rm F}\gsim3
$ (e.g., Ellis et al. 1997). Hereafter, we call such models with ($\tau_
{\rm SF}$, $t_{\rm gw}$, $z_{\rm F})=(0.1$ Gyr, 0.353 Gyr, $z_{\rm F}\geq3
$) the single-burst models. For the initial mass function, we assume a
power-law mass spectrum with a slope of $x=1.10$ in the range of $
0.1M_\odot\leq M\leq60M_\odot$. We examine whether the single-burst models
can reproduce both observed redshift distributions and number
counts. Fig.\ref{fig:nz_singleburst}(a) compares the observed
redshift distribution of the $K$-band sample (solid histogram) with
predictions by a series of single-burst models with different $z_{\rm F}$. Four
lines correspond to $z_{\rm F}=3,$ 5, 8 and $\infty$. It is found that
the single-burst models overpredict the number of galaxies at $z\gsim1.4
$ irrespective of $z_{\rm F}$. In particular, any single-burst model predicts a
sharp peak near $z_{\rm F}$, because galaxies just after their formation epoch
are very bright. Such a peak would disappear if we assume that the formation
redshift of early-type galaxies is not a single value but is distributed
uniformly in time between $z_{\rm F}=3$ and $\infty$. However, the
overprediction at $z\gsim1.4$ still remains, and these models are not
consistent with the observations of redshift distribution. We also compare
predictions with the observed redshift distribution of $I$-band sample, and
find similar overpredictions. The models shown
Fig.\ref{fig:nz_singleburst}(a) overpredict the $K$-band counts, but they are
marginally consistent with the $B$-band and $I$-band counts. However, they are
ruled out by the redshift distributions anyway.

\subsection{Models with interstellar dust}

\begin{figure}
\vspace{-1.5cm}
\begin{center}
  \leavevmode
  \epsfxsize 1.0\hsize
  \epsffile{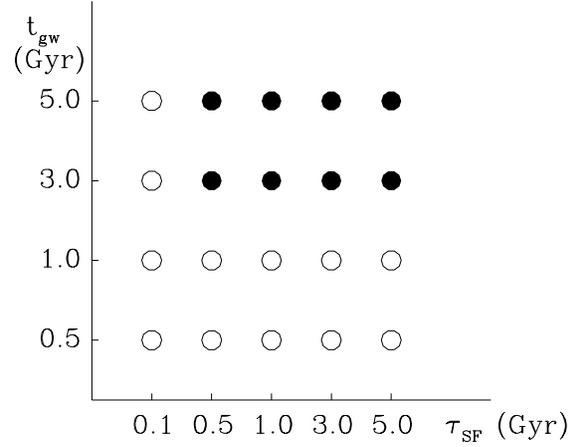}
\end{center}
\vspace{-0.8cm}
\caption{The
filled circles indicate a combination of parameters with which the model
reproduces observed number counts and redshift distributions for appropriate
choice of $z_{\rm F}$ and $E(B-V)_p$, while open circles show the parameters
which do not account for the observations.
\label{fig:result}}
\end{figure}

\begin{figure}
\begin{center}
  \leavevmode
  \epsfxsize 0.9\hsize
  \epsffile{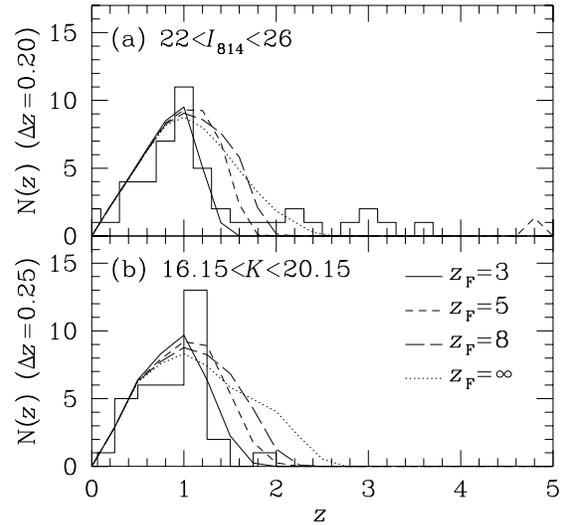}
\end{center}
\vspace{-0.5cm}
\caption{Redshift distributions (histogram) of the $I$-band sample (a) and
the $K$-band sample (b). Predictions of models using $\tau_{\rm SF}=0.5
$ Gyr, $t_{\rm gw}=3.0$ Gyr and $E(B-V)_{\rm p}=2$ are shown by lines. Four
lines correspond to $z_{\rm F}=3,$ 5, 8 and $\infty$.
\label{fig:nz_t05tw30}}
\end{figure}

\begin{figure*}
\vspace{-5.5cm}
\begin{center}
  \leavevmode
  \epsfxsize 1.0\hsize
  \epsffile{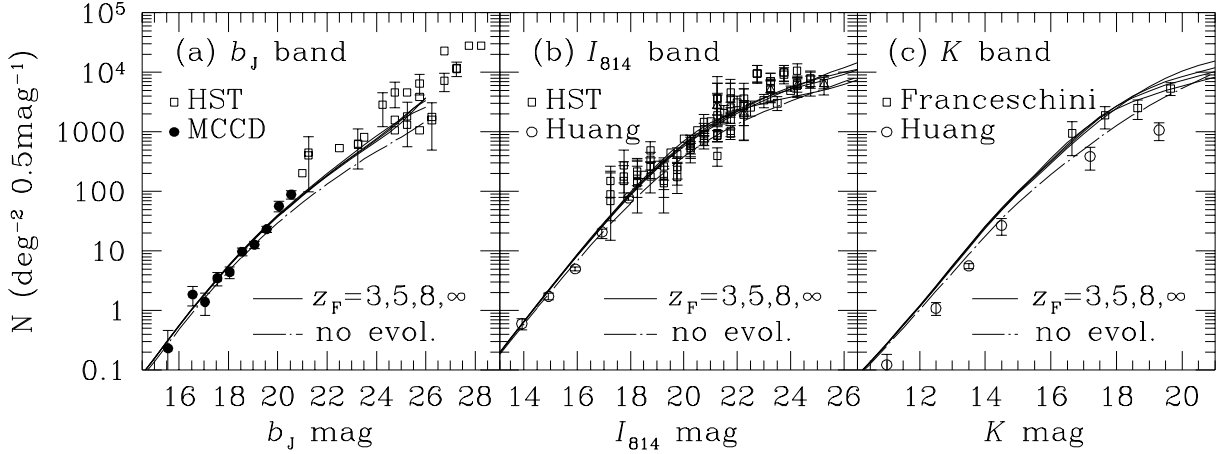}
\end{center}
\vspace{-6cm}
\caption{Differential number counts as a function of apparent
magnitude. Panels (a), (b) and (c) are for $B$, $I$ and $K
$ band, respectively. Predictions of models using $\tau_{\rm SF}=0.5
$ Gyr, $t_{\rm gw}=3.0$ Gyr and $E(B-V)_{\rm p}=2$ are shown by
lines. Four lines correspond to $z_{\rm F}=3,$ 5, 8 and $\infty$. Dash-dotted
lines are for no evolution models. See
the text (\S2) for data sources. MCCD data are used to normalise the local
luminosity function. We do not calculate the counts beyond $B>26$ mag because
of the lack of reliable $K$-correction for redshift above $z=2$.
\label{fig:nm_t05tw30}}
\end{figure*}

In order to suppress the overprediction found for the single-burst model, we
introduce the effect of dust extinction. The paucity of galaxies at $z\gsim1.4
$ may not be due to dust extinction but to some observational selection
effects. The dimming of surface brightness by a factor of $(1+z)^4$, for
example, would decrease the detection rate of early-type galaxies at high
redshifts. To investigate this effect, we examine the evolution of surface
brightness and effective radius of the normal early-type galaxies seen in
the local universe on the basis of the single-burst models, and find that
most of these galaxies up to $z\sim z_{\rm F}$ survive the selection criteria
for surface brightness and radius  set for Franceschini et
al.'s (1998) sample, if they are brighter than $K=20.15$ mag, the limiting
magnitude of Franceschini et al.'s (1998) sample. This result is also
supported by a similar test using the early-type galaxies of Franceschini et
al.'s (1998) sample for which surface brightness, effective radius and
redshift are given. Accordingly, the surface brightness selection is not
a problem. Another possible selection effect comes from type
classification. Early-type galaxies undergoing intensive star
formation, where a significant amount of gas
exists, might be classified as late-type galaxies. Such selection effects
need further investigation. In this paper, we assume that such selection
effects are negligible.

Following Vansevi\v{c}ius, Arimoto \& Kodaira (1997), we express the $B
$-band optical depth of a galaxy as

\begin{equation}
\tau(B)=Cf_{\rm g}Z_{\rm g} \label{eq1}
\end{equation}

\noindent
where $f_{\rm g}$ is the gas fraction of the galaxy, $Z_{\rm g}$ is the
metallicity of the galaxy, and $C$ is a constant. The internal
extinction in the $B$ band is $A_{\rm B}=-2.5\log[\exp\{-\tau(B)\}]
$ mags. We adopt the extinction curve given in
Cardelli, Clayton \& Mathis (1989). Then, the colour excess is
expressed as $E(B-V)\simeq0.245\,\tau(B)$. Metallicity $Z_{\rm g}$ increases
with time, while $f_{\rm g}$ is a decreasing function of time, with $
f_{\rm g}=0$ at $t\ge t_{\rm gw}$. Therefore, $\tau(B)$ (and thus $E(B-V)
$) has a peak value. We treat the peak value, $E(B-V)_{\rm p}$, as a free
parameter and change it from $E(B-V)_{\rm p}=0$ to 2. It should be
noted, however, that the nominal peak value obtained from eq.(\ref{eq1}) is not
necessarily reached at $t\le t_{\rm gw}$. Actually, $E(B-V)=0$ for $t\geq
t_{\rm gw}$, of course. However, it is the nominal peak value that we change. $
E(B-V)_{\rm p}=2$ corresponds to extraordinary strong extinction.

Fig.\ref{fig:nz_singleburst}(b) compares the observed redshift
distribution of $K$-band sample (solid histogram) with predictions by
models for $E(B-V)_{\rm p}=2$ with different $z_{\rm F}$. It is found that
there still remains the overprediction. This is because the extinction, though
very strong, works only in a very short period ($\lsim0.3$ Gyr) due to the
small values for $\tau_{\rm SF}$ and $t_{\rm gw}$. The models shown
Fig.\ref{fig:nz_singleburst}(b) are more or less consistent with the number
counts. However, they are also ruled out by the redshift distributions.

If larger values are adopted for $\tau_{\rm SF}$ and $t_{\rm gw}$, the
duration when extinction works becomes longer. In order to investigate this
effect, we change $\tau_{\rm SF}$ and $t_{\rm gw}$ as free parameters in $
0.1\leq\tau_{\rm SF}\leq5.0$ Gyr and $t_{\rm gw}\geq0.5
$ Gyr, respectively, and examine whether there exists a combination of $
\tau_{\rm SF}$ and $t_{\rm gw}$ that does not
overpredict galaxies at $z\gsim1.4$ and is consistent with observed number
counts. We regard a model to be consistent with
the observations if the model reproduces both observed number counts and
redshift distributions for appropriate choices of $z_{\rm F}$ and $
E(B-V)_{\rm p}$. Results are shown in Fig.\ref{fig:result}, where the filled
circles indicate a combination of parameters which is consistent with the
observations while open circles show the parameters which do not account
for the observations. We confirm that the combinations indicated by filled
circles give $B-V$ colours at $z=0$ which are consistent with that of local
early-type galaxies ($0.8<B-V<1.1
$; Fukugita, Shimasaku \& Ichikawa 1995). As an example, we present the
predicted redshift-distributions using $\tau_{\rm SF}=0.5$ Gyr, $
t_{\rm gw}=3.0$ Gyr and $E(B-V)_{\rm p}=2$ in Fig.\ref{fig:nz_t05tw30}. We
regard the cases when $z_{\rm F}=3$ in Fig.\ref{fig:nz_t05tw30} as being
consistent with the observation. This combination is also consistent with the
observed number counts as shown in
Fig.\ref{fig:nm_t05tw30}. In Fig.\ref{fig:nz_t05tw30}(a), there are several
galaxies at $z\gsim1.5$ in the $I$ band sample, while the models predict no
galaxies there. We do not think, however, that this discrepancy is
significant. It is plausible that the amount of dust extinction of actual
galaxies has various values, unlike our models that assume the same dust
extinction for all galaxies. A small fraction of weaker extinction galaxies
could explain those galaxies at $z\gsim1.5$.

We find from Fig.\ref{fig:result} that if $\tau_{\rm SF}\geq0.5$ Gyr and $
t_{\rm gw}>1.0$ Gyr are taken, the predicted number counts and redshift
distributions are consistent with the observations. However, $
E(B-V)_{\rm p}\geq1$ is needed for all cases. This value is much larger
than the colour excess of Lyman break galaxies at $z\simeq3$, $E(B-V)\sim0.3
$, found by Sawicki \& Yee (1998). In Fig.\ref{fig:result}, we show the
result only for the case of $t_{\rm gw}\leq5.0$ Gyr. In fact, we investigate
the case that galactic wind does not blow, and find that the predicted
number counts and redshift distributions are consistent with the observations
as well. However, colours of these galaxies at $z=0$ are inconsistent with
observations. 

It is interesting to examine whether or not the models indicated as the filled
circles in Fig.\ref{fig:result} are consistent with the observations of
early-type galaxies in clusters as well. We find that these models are
marginally consistent with the observed evolution of luminosity and mean
colours of cluster early-type galaxies up to $z=1.27$, the redshift of the
most distant cluster at present. (We cannot discuss the slope and scatter
of the colour-magnitude relations of cluster early-type galaxies, because we
assume here that all early-type galaxies have the same age and
metallicity.) At $z>1.27$, however, these models predict quite different
evolution of luminosity and colours from that based on the single-burst
models, due to the effect of dust extinction.

As seen in Fig.\ref{fig:nm_t05tw30}(c), $K$-band number
counts of Huang et al. (1998) at $17\lsim K\lsim19$ are lower than
those of Franceschini et al. (1998) by more than a factor of three. The reason
for this discrepancy is not clear for us. However, $K$-band number counts
of Huang et al. (1998) in bright magnitude, $K\lsim14$, where the
effect of galaxy evolution is negligible, are also lower than the
prediction, though the predictions for $B$ and $I$ counts at bright
magnitudes are in good agreement among different observations. Because 
we cannot reproduce Huang et al.'s (1998) data for $K$-band counts, we ignore
them in comparing with models.

\section{CONCLUSION AND DISCUSSION}

We investigate pure luminosity evolution models for early-type galaxies
which do not exhibit either number density change or morphology
transition, and examine whether these models are consistent with observed
number counts and redshift distributions. We summarize our findings as follows:

\begin{enumerate}

\item We find that single-burst models are inconsistent with observed
redshift distributions irrespective of $z_{\rm F}$ (if $z_{\rm F}>3$), due
to the overpredictions of galaxies at $z\gsim1.4$, even if dust extinction
which is in effect during star formation is taken into account.

\item In order for dust extinction to be more effective, we change $
\tau_{\rm SF}$ and $t_{\rm gw}$ as free parameters over the ranges of $
0.1\leq\tau_{\rm SF}\leq5.0$ Gyr and $t_{\rm gw}\geq0.5
$ Gyr, respectively, and examine whether there exists
a combination of $\tau_{\rm SF}$ and $t_{\rm gw}$ that does not overpredict
galaxies at $z\gsim1.4$ and is also consistent with observed number counts. We
find that models with $\tau_{\rm SF}\geq0.5$ Gyr and $t_{\rm gw}>1.0$ Gyr can
be made consistent with the observed redshift distributions and number
counts, if we introduce strong extinction ($E(B-V)_{\rm p}\ge1$).

\end{enumerate}

Our finding that the single burst models cannot express the evolution of
field early-type galaxies well is not new. Franceschini et
al. (1998) analysed $K$-band number counts and redshift
distribution for field early-type galaxies, and found that models which
assume the duration of star formation to be about 3 Gyr with dust
extinction during star formation are preferred. Our study not only confirms
their results but also extends their study: using $B$-, $I$- and $K$-band
number counts and $I$- and $K$-band redshift distributions, we systematically
study what combinations of $\tau_{\rm SF}$ and $t_{\rm gw}$ can reproduce the
observation. He \& Zhang (1999) presented a redshift distribution for galaxies
limited to $22.5<b_{\rm j}<24.0$ and $B-K>5.5$, which they considered to be
early-type galaxies, and compared it with the predictions of single-burst
models and models with $\tau_{\rm SF}=1$ Gyr. They found that these models
overpredict the observed redshift distribution at $z>0.8$, and suggested that
number evolution may be essential for early-type galaxies. However, we
confirm that using our models with strong dust extinction, the
overpredictions of the redshift distribution can be removed. Driver
et al. (1998) analysed $I$-band number counts and redshift distribution for
early-type galaxies and
found that the single-burst models overpredict galaxies in redshift
distribution, which is consistent with our results. Im et
al. (1999) analysed $I$-band redshift distribution for early-type
galaxies and found that observations are consistent with those expected
from passive luminosity evolution or are only in slight disagreement with
the nonevolving model. However, their results are not in conflict with
ours, because the redshift distributions they analysed are
at $I<21$, where most galaxies are at $z<1.0$. 

It is true that the pure luminosity evolution models with dust extinction are
not the only model which reproduces the observed number counts and redshift
distributions. Models with number density change or morphology
transition, such as those based on hierarchical clustering
scenarios (e.g., Kauffmann, White \& Guiderdoni 1993; Baugh, Cole \& Frenk 1996) may
account for the observations. However, we can at least conclude that early-type
galaxies in the field environment have a different evolution history from
the single-burst models, if the observational selection effects are not
significant.

\bigskip

We would like to thank James Annis for useful comments. This work is
supported in part by
Grants-in-Aid (07CE2002,11640228,10440062) from the
Ministry of Education, Science, Sports and Culture of Japan. W.K. acknowledges
the travel support by the Hayakawa Fund of the Astronomical Society of Japan.

\end{document}